\documentclass[aps,reprint,amsmath,amssymb,superscriptaddress]{revtex4-1}

\usepackage{graphicx}

\usepackage[usenames, dvipsnames]{xcolor}

\newcommand{\bra}[1]{\left< #1 \right|}
\newcommand{\ket}[1]{\left| #1 \right>}
\newcommand{\melement}[3]{\left< #1 \right| #2 \left| #3 \right>}
\newcommand{\vect}[1]{\mathbf{#1}}

\begin{document}
\title{The mesoscopic magnetron as an open quantum system}

\author{Tadeusz Pudlik}
\affiliation{Department of Physics, Boston University, Boston, MA 02215, USA}
\author{Antonio H.~Castro Neto}
\affiliation{Centre for Advanced 2D Materials and Graphene Research Centre, National University of Singapore, 2 Science Drive 3, Singapore, 117542}
\author{David K.~Campbell}
\email[To whom correspondence should be addressed, at ]{dkcampbe@bu.edu}
\affiliation{Department of Physics, Boston University, Boston, MA 02215, USA}

\date{\today}

\begin{abstract}
Motivated by the emergence of materials with mean free paths on the order of microns, we propose a novel class of solid state radiation sources based on reimplementing classical vacuum tube designs in semiconductors.  Using materials with small effective masses, these devices should be able to access the terahertz range.  We analyze the DC and AC operation of the simplest such device, the cylindrical diode magnetron, using effective quantum models.  By treating the magnetron as an open quantum system, we show that it continues to operate as a radiation source even if its diameter is only a few tens of magnetic lengths.
\end{abstract}
\maketitle

Magnetrons are vacuum tubes that convert a DC voltage into electromagnetic radiation in the microwave range.  Historically, the designs of vacuum tube and solid state radiation sources were radically different, as transport in semiconductors was limited to the diffusive regime.  Today, as a wide range of two-dimensional materials transition from basic research into the toolkit of device designers, it is becoming possible to build solid state devices characterized by ballistic transport~\cite{Liang2007,Du2008b, Mayorov2011a, Wang2015}.  Thus, a broad range of vacuum tube designs perfected over the decades---magnetrons, crossed-field amplifiers, gyrotrons, etc.~\cite{Gilmour2011}---can serve as direct inspiration for a new generation of solid state radiation sources.  Such devices could retain some of the advantages of tubes, such as their wide frequency tunability, without the disadvantages of cost and weight associated with vacuum technology.  But the new solid state devices would be different in one critical respect: due to their small size and the presence of band structure, they will exhibit quantum effects.  Past work on these devices has focused on analogs of linear beam tubes~\cite{Gribnikov2003,Asada2003,Ryzhii2009}. Here, we discuss crossed-field designs.  As a first step in the investigation of this class of devices, we propose a simple quantum model of the magnetron.

We focus on the most basic magnetron design, the so-called cylindrical anode or Hull magnetron~\cite{Hull1928, Collins1964, Ma2004}.  We briefly review the classical mechanism of its operation before developing a fully quantum model in which both the electron motion and the electromagnetic field are quantized.  Our model shows that net radiation gain persists deep into the quantum regime and that startup will take place even if the field initially contains no photons, thanks to spontaneous emission.  Most importantly, though, our work provides a framework for designing solid-state analogs of the magnetron and related devices.

\section{The classical model of the magnetron}
\label{sec:classical_model}

\begin{figure}
\begin{center}
\includegraphics[width=\columnwidth]{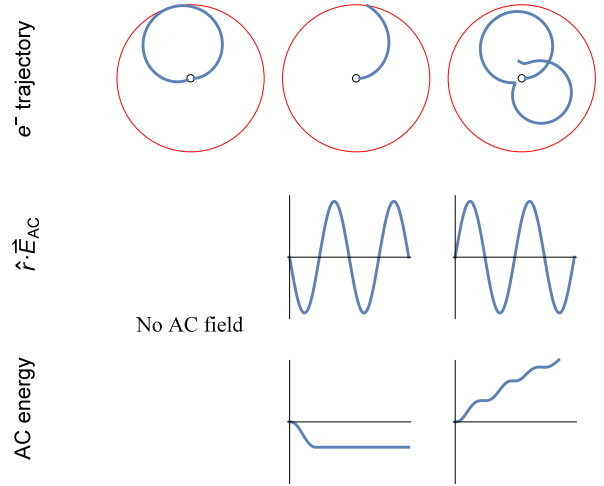}
\caption{Principle of operation of the cyclotron resonance magnetron.  The DC fields are set up so that $V \approx V_H$ (first column).  If an electron is emitted when the AC voltage has the same polarity as the DC, the electron collides with the anode before removing much energy from the field (second column).  If it is emitted when the AC voltage has a polarity opposite to the DC, it remains in the device for a longer time and exchanges more energy with the field (third column).  See Appendix~\ref{sec:classical_model_derivation} for the details of the model used to generate these plots.
\label{fig:magnetron_schematic}}
\end{center}
\end{figure}

The cyclotron resonance magnetron consists of two coaxial conducting cylinders; see the top row of Figure~\ref{fig:magnetron_schematic}.  The inner cylinder is kept at a negative potential and constitutes the cathode, the grounded outer cylinder is the anode, and the space between them is evacuated.  An external DC magnetic field points along the axis of the cylinders.

An electron emitted by the cathode performs cyclotron motion within the device.  The radius of the cyclotron orbit increases with the accelerating voltage, $V$.  Below the so-called Hull cutoff voltage, $V_H$, the diameter of the orbit is smaller than the device radius, and the emitted electrons never reach the anode.  Above $V_H$, the electrons reach the anode and the device is conducting.

Now, consider a device operating just below $V_H$.  Connect a resonant circuit tuned to the cyclotron frequency to the cathode and anode; this generates an AC voltage in addition to the DC one.  Those electrons emitted when the AC voltage has the same polarity as the DC will absorb energy from the electromagnetic field---but since they are accelerated by a voltage $V > V_H$, they will be removed from the device by a collision with the anode during their first orbit (see first column of Figure~\ref{fig:magnetron_schematic}).  Those electrons emitted when the AC voltage has the opposite polarity lose energy to the field and remain in the device.  Crucially, by the time these slowed electrons reach the apex of their orbit and turn around, the polarity of the AC voltage reverses, so that they are once again giving up energy to the field.  Thus, those electrons that are not quickly removed by a collision with the anode continuously transfer their energy to the electromagnetic field (see third column of Figure~\ref{fig:magnetron_schematic}).  The result is net emission.

An important subtlety is that the interaction with the electromagnetic field perturbs the electron's orbit, leading to a gradual change of the relative phase of the electron's and the field's oscillations.  Therefore, even the electrons which initially contribute energy to the field will eventually absorb it instead. From an energy perspective, the problem can be stated as follows: if the only way for the electron to be removed from the device is a collision with the anode, then by the time the electron is removed it must have absorbed energy on net from the AC field.  To eliminate this problem, \emph{all} electrons are removed from the device on some timescale long compared to the cyclotron frequency but short relative to the dephasing time, even if they are in orbits too small to reach the anode.  In vacuum magnetrons, this can be achieved by tilting the magnetic field slightly away from the electrodes' axis.

\section{Proposed device}
\label{sec:proposed_device}

The classical model of the previous section suggests the device design shown in Figure~\ref{fig:proposed_device}.  As reviewed in Appendix~\ref{sec:classical_model_derivation}, in the cylindrical geometry the Hull voltage is,
\begin{equation}
\frac{V_H}{B^2} = \frac{e s_a^2}{8\mu} \left(1 - \frac{s_c^2}{s_a^2}\right)^2,
\end{equation}
where $s_c$ is the cathode (inner) diameter and $s_a$ the anode (outer) diameter, $e$ is the carrier charge, $\mu$ the carrier mass, and $B$ the DC magnetic field, related to the frequency of operation through the cyclotron condition,
\begin{equation}
\omega = \frac{eB}{\mu}.
\end{equation}
Critically, unlike in the vacuum device, the effective charge carrier mass can be controlled by appropriate choice of material.  This allows access to higher emission frequencies.  For example, in a monolayer of GaSe, with an effective mass of $0.053\,m_e$~\cite{Wickramaratne2015}, an emission frequency of 1 THz should be achieved at a field of 1.9 Tesla. (The device geometry and the Hull condition set the voltage drop at 1.6 V.)  Other two-dimensional materials with small effective masses and parabolic band structures could be used as well.  Bilayer graphene would be a natural candidate, but its band structure shows deviations from parabolicity, and consequently its Landau levels are only approximately uniformly spaced~\cite{Pereira2007}.  Another possibility would be to use the two-dimensional electron gas in a AlGaAs/GaAs or AlGaAs/InGaAs/GaAs heterostructure, with effective masses of $0.068\,m_e$~\cite{Zudov2001} and $0.073\,m_e$~\cite{Liu1988}, respectively.  This would require a device geometry slightly different from that shown in Figure~\ref{fig:proposed_device}, with electrodes penetrating capping layers to contact the two-dimensional electron gas.

\begin{figure}
\begin{center}
\includegraphics[width=0.7\columnwidth]{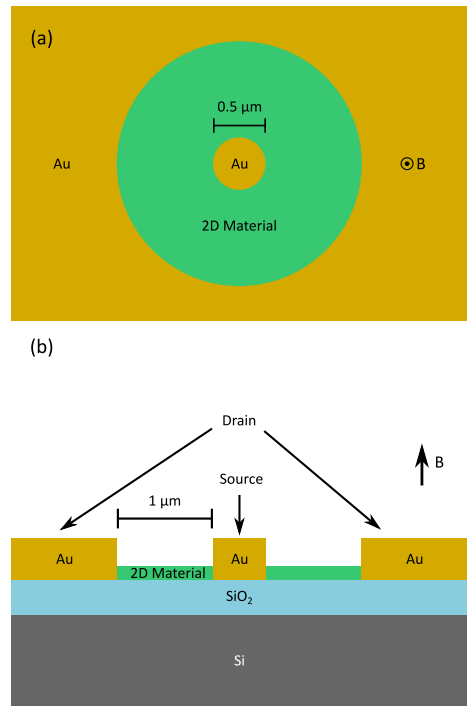}
\caption{Solid state magnetron.  \emph{(a)} Top view.  \emph{(b)} Side view.
\label{fig:proposed_device}}
\end{center}
\end{figure}

Unfortunately, the validity of the simple classical model is far from obvious: the distance between the electrodes is less than 50 magnetic lengths ($\sqrt{\hbar/eB}$), a scale at which the wave nature of the electron cannot be ignored. In the remainder of this paper, we propose and develop a fully quantum model of a solid state magnetron.

\section{Quantum model: DC operation}

In the classical picture described in Section~\ref{sec:classical_model}, a DC magnetic field and an absorbing boundary allow for the transfer of energy from a DC voltage source to an AC signal.  We will now describe the same process from a quantum perspective.

\begin{figure}
\begin{center}
\includegraphics[width=0.5\columnwidth]{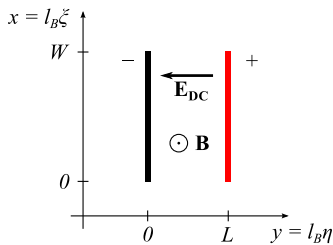}
\caption{A schematic of the quantum magnetron model.  We will assume $W \gg L$.
\label{fig:rectangular_schematic}}
\end{center}
\end{figure}

To simplify the analysis, we will discuss a rectangular, rather than cylindrical, geometry as shown in Figure~\ref{fig:rectangular_schematic}. (The rectangular geometry is simpler because the DC electric field between the electrodes has a constant magnitude.  The general mechanism of device operation is unchanged.)  Furthermore, we will restrict our considerations to a planar, or 2D, device.  Within the device region ($y \in [0, L]$) there are constant crossed electric and magnetic fields $\vect{E}$ and $\vect{B}$, while outside of it---in the electrodes---the fields are zero.  The motion of an electron subject to these external potentials is described by the Hamiltonian,
\begin{equation}
\label{eq:hamiltonian}
H = \frac{1}{2\mu} (\vect{p} + e\vect{A})^2 - e \Phi,
\end{equation}
where $-e < 0$ is the electron charge and $\mu$ the electron mass.

We choose the Landau gauge, in which
\begin{equation}
\Phi = \begin{cases}
0 & \text{for $y < 0$,}\\
E y & \text{for $y \in [0, L]$,}\\
E L & \text{for $y > L$,}
\end{cases}
\quad
\vect{A} = \begin{cases}
0   & \text{for $y < 0$,}\\
B y & \text{for $y \in [0, L]$,}\\
B L & \text{for $y > L$.}
\end{cases}
\end{equation}
By introducing the cyclotron frequency and magnetic length,
\begin{equation}
\omega_c = \frac{eB}{\mu}, \quad l_B = \sqrt{\frac{\hbar}{eB}},
\end{equation}
we can rewrite the Hamiltonian in terms of dimensionless variables,
\begin{equation}
\xi = x/l_B,\quad \eta = y/l_B,\quad \alpha = \frac{e E l_B}{\hbar\omega_c}, \quad \Lambda = L/l_B,
\end{equation}
as,
\begin{equation}
H = \frac{1}{2}\hbar\omega_c\times \begin{cases}
p_\xi^2 + p_\eta^2 & \text{for $\eta < 0$,}\\
(p_\xi + \eta)^2 + p_\eta^2 - 2\alpha \eta & \text{for $\eta \in [0, \Lambda]$,}\\
(p_\xi + \Lambda)^2 + p_\eta^2 - 2\alpha \Lambda & \text{for $\eta > \Lambda$.}
\end{cases}
\end{equation}
In this gauge, $\xi$ does not appear in the Hamiltonian and $p_\xi$ is a constant of the motion.  The energy eigenstates $\{\psi\}$ can be taken to be simultaneous eigenstates of $p_\xi$:
\begin{equation}
\psi_k(\xi, \eta) = e^{\imath k \xi} \phi_k(\eta),
\end{equation}
where $\phi_k$ is an eigenstate of the one-dimensional Hamiltonian,
\begin{equation}
H_k = \frac{1}{2}\hbar\omega_c \left(p_\eta^2 + V_k(\eta)\right).
\end{equation}
The effective potential $V_k(\eta)$ is,
\begin{equation}
V_k(\eta) = \begin{cases}
k^2 & \text{for $\eta < 0$,}\\
(k + \eta)^2 - 2\alpha \eta & \text{for $\eta \in [0, \Lambda]$,}\\
(k + \Lambda)^2 - 2\alpha \Lambda & \text{for $\eta > \Lambda$.}
\end{cases}
\end{equation}
If we neglected the electrodes (assumed the device region extends from $-\infty$ to $\infty$, rather than from 0 to $\Lambda$), the effective potential would be parabolic, leading to eigenstates and energies of a simple harmonic oscillator,
\begin{gather}
\psi_{m, k} = \sqrt{\frac{l_B}{W}} e^{\imath k \xi} e^{-(\eta + k - \alpha)^2} \mathcal{H}_m(\eta + k - \alpha),\\
\epsilon_m = \hbar \omega_c \left(m + \frac{1}{2}\right) + \hbar \omega_c \left(\alpha k - \frac{\alpha^2}{2}\right),
\end{gather}
where $\mathcal{H}_m$ is the $m$'th Hermite polynomial, $m = 0,\, 1,\, 2,\, \ldots$; $k = \frac{2\pi l_B}{W}p$ for $p \in \mathbb{Z}$; and $W$ is the device width.  If the device region is large but finite, we expect the eigenstates to take a similar form within this region~\cite{Halperin1982}.  This implies we are interested in states the center of which is within the device, or those for which $k \in (\alpha - \Lambda, \alpha)$.

The parameter $\alpha$ is a dimensionless measure  of the electric field strength.  If the device is operated at the Hull cutoff voltage, i.e. if the width of the classical cycloid trajectory is equal to the device length ($\frac{2 E}{B\omega_c} = L$), then
\begin{equation}
\alpha = \frac{\Lambda}{2},
\end{equation}
the allowed $k$ values are $k \in (-\Lambda/2, \Lambda/2)$ and the effective potential takes the simple form,
\begin{equation}
\label{eq:effective_potential}
V_k(\eta) = \begin{cases}
k^2 & \text{for $\eta < 0$,}\\
(k + \eta)^2 - \Lambda \eta & \text{for $\eta \in [0, \Lambda]$,}\\
k^2 + 2 k \Lambda & \text{for $\eta > \Lambda$.}
\end{cases}
\end{equation}
Note that the potential is parabolic within the device region and always contains a bound state as well as higher-energy scattering states.  In what follows we will restrict our attention to the $k = 0$ case, corresponding to an electron injected into the device with no momentum in the $x$ (or $\eta$) direction.

\begin{figure}
\centering
\includegraphics[width=\columnwidth]{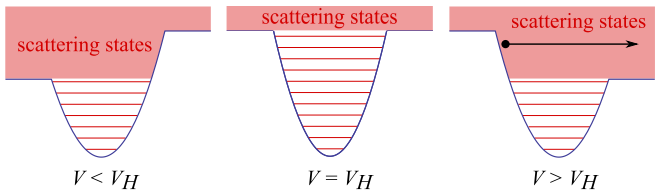}
\caption{In the absence of an AC field, the magnetron operates as a diode.  The effective potential $V_0(\eta)$ is plotted above for three values of $\alpha$ (or, equivalently, DC voltage), for $k=0$.  Below and at the Hull voltage $V_H$, no conduction is observed, as an electron localized near the left edge of the device can be decomposed into bound states.   Above $V_H$, the electron can only be decomposed into scattering states, and the current is a constant independent of voltage.
\label{fig:quantum_diode}}
\end{figure}

In the absence of an AC field, the magnetron operates as a diode (see Figure~\ref{fig:quantum_diode}).  For $\alpha \leq \frac{\Lambda}{2}$, an electron initially localized just within the device, by the cathode, can be decomposed into eigenstates that do not enter the anode; consequently, there is no current.  For $\alpha > \frac{\Lambda}{2}$, however, states localized near the cathode can be decomposed into scattering states extending to the anode, and current is observed.  Increasing $\alpha$ even further affects states initially localized closer to the device's center, but since the electrons enter the device near the cathode, this does not increase the current.  Thus, the current-voltage characteristic is approximately a step function, in agreement with the classical model. 

\section{Quantum model: AC operation}

\begin{figure}
\begin{center}
\includegraphics[width=0.8\columnwidth]{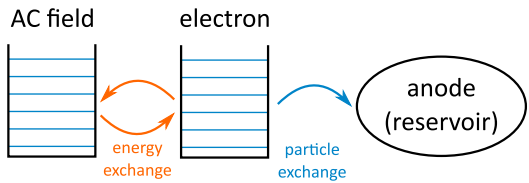}
\caption{A schematic of the two-mode quantum model of magnetron AC operation.
\label{fig:dimer_schematic}}
\end{center}
\end{figure}

To treat the interaction of the electron with the AC field, we will introduce a simplified effective model schematically depicted in Figure~\ref{fig:dimer_schematic}.  We will consider only one mode of the AC field, described by the Hamiltonian,
\begin{equation}
H_\mathrm{field} = \hbar \omega \left(\hat{b}^\dagger \hat{b} + \frac{1}{2}\right),\quad [\hat{b}, \hat{b}^\dagger] = 1.
\end{equation}
Since the effective potential $V_0(\eta)$ within the device region is parabolic, we will approximate the single-particle electron energy with another harmonic mode:
\begin{equation}
\label{eq:eff_hamiltonian_electron}
H_\mathrm{electron} = \hbar \omega \left(\hat{a}^\dagger \hat{a} + \frac{1}{2}\right), \quad [\hat{a}, \hat{a}^\dagger] = 1.
\end{equation}
The interaction energy between the electron and the field is~\cite[p.~57]{Marcuse1980},
\begin{equation}
\begin{split}
H_\mathrm{int} &= \frac{\hbar e}{2 L} \frac{1}{\sqrt{\mu C}}(\hat{a}^\dagger + \hat{a})(\hat{b}^\dagger + \hat{b})\\
 &= \frac{e}{2\Lambda} \sqrt{\frac{\hbar \omega}{C}} (\hat{a}^\dagger + \hat{a})(\hat{b}^\dagger + \hat{b}),
\end{split}
\end{equation}
where $C$ is the magnetron's capacitance.  Defining
\begin{equation}
J = \frac{e}{2\Lambda}\frac{1}{\sqrt{\hbar\omega C}},
\end{equation}
and performing a rotating wave approximation---justified as long as the coupling is small, $J \ll \omega$~\cite{Walls2008}---we can write the total effective Hamiltonian in dimensionless form,
\begin{equation}
\label{eq:eff_hamiltonian_total}
H_\mathrm{eff} = \hbar \omega \left(\hat{a}^\dagger \hat{a} + \hat{b}^\dagger \hat{b} + J(\hat{a}^\dagger \hat{b} + \hat{b}^\dagger \hat{a}) + 1\right).
\end{equation}
This model incorporates the effects of the DC and AC fields but does not describe the electron being absorbed by the anode.  In the absence of this dissipative process, neither the classical nor the quantum model predicts net energy transfer to the AC field. 

To model electron loss from the magnetron cavity, we couple the system described by $H_\mathrm{eff}$ to a fermionic reservoir representing the anode:
\begin{equation}
H_\mathrm{E} = \sum_k \varepsilon_k \hat{r}_k^\dagger \hat{r}_k,\quad \{\hat{r}_k, \hat{r}_l^\dagger\} = \delta_{k,l}.
\end{equation}
The annihilation (creation) of an electron in a state $n$ of the approximate harmonic potential within the device is described by the fermionic operator $\hat{c}_n$ ($\hat{c}_n^\dagger$), with $\{\hat{c}_m, \, \hat{c}_n^\dagger\} = \delta_{m,n}$.  The electron part of the Hamiltonian can still be written as Eq.~(\ref{eq:eff_hamiltonian_electron}),\footnote{Except that the zero-point energy is no longer $1$ but rather $\frac{1}{2} + \sum_n \hat{c}_n^\dagger \hat{c}_n$.} but the operator $\hat{a}$ is now \emph{defined} as,
\begin{equation}
\hat{a} = \sum_{n=0}^\infty \sqrt{n+1} \,\hat{c}_n^\dagger \hat{c}_{n+1}.
\end{equation}
On the subspace of one-electron states, the operator $\hat{a}$ satisfies the usual bosonic commutation relation.  The most general form of the coupling between the electron in the device and the fermionic reservoir is,
\begin{equation}
\label{eq:coupling}
\hat{V} = \sum_n \sum_k \gamma_{n, k} (\hat{c}_n + \hat{c}_n^\dagger)(\hat{r}_k + \hat{r}_k^\dagger).
\end{equation}
Since we are not interested in the dynamics of the reservoir, we will treat the electron and AC mode as an open quantum system~\cite{Breuer2002, Wiseman2010}.  This will allow us to derive equations of motion for the system degrees of freedom only.

The details of the derivation, which uses the approach of Tomka~\cite{Tomka2014} originally developed by Beaudoin et al.~\cite{Beaudoin2011}, are in Appendix~\ref{sec:master_eq_derivation}.  Here, we will only recapitulate the assumptions:
\begin{enumerate}
\item Born approximation: the density matrix of the anode is only negligibly affected by the interaction with the electron in the device.
\item Markov approximation: the anode correlation functions decay at a rate much faster than any other timescale of the model.
\item Rotating wave approximation in the system-bath coupling.
\item The anode is in thermal equilibrium, and the Fermi factor of the relevant levels is approximately zero.  This implies the anode never emits electrons into the device, and any electron impinging on the anode from the device will be absorbed.
\item We ignore the frequency shift of the device energy levels that results from the coupling to the reservoir.    The frequency shift is in general not negligible.  However, its main consequence is that the actual emission frequency of the device is different from the DC cyclotron frequency $eB/\mu$.
\item The system-reservoir coupling constants satisfy $\gamma_{n,k} = \gamma_{n, k'} \equiv \gamma_n$ for all $n, k, k'$.  This is a technical assumption which simplifies the form of the final results; it could be substantially relaxed if we wished to make more specific assumptions about the band structure of the anode.
\end{enumerate}
Given these assumptions, the time evolution of the system density matrix is given by an equation of the Lindblad form,
\begin{equation}
\label{eq:lindblad}
\dot{\rho}  =-\imath [H_\mathrm{eff}, \rho] + \hat{A}\rho\hat{A} - \frac{1}{2}\left(\hat{A}^\dagger \hat{A} \rho + \rho \hat{A}^\dagger \hat{A}\right),
\end{equation}
where the operator $\hat{A}$ is,
\begin{equation}
\label{eq:lindblad_operator}
\hat{A} = \sqrt{2\pi \sigma} \sum_{n = 0}^\infty \gamma_n \hat{c}_n,
\end{equation}
with $\sigma$ the anode density of states.\footnote{The density of states is a constant because we have assumed the device, including the electrodes, to be confined to a plane.} Thus, the effect of the anode is to remove electrons from level $n$ at a rate $\gamma_n$.

What are the values of the dissipation rates?  The scattering modes of the effective potential $V(\eta)$ [Eq.~(\ref{eq:effective_potential})] overlap with the electrodes; an electron excited into one of these levels will be removed from the device at a rate of order $L/v \sim L\sqrt{\frac{\mu}{\hbar \omega}} \gg J$.  But as we observed while discussing the classical device, the lower energy electrons must also be removed from the device, albeit at a slower rate $\sim J$, if net emission is to be observed.  Therefore, we will assume $\gamma_n$ is a step function of $n$, taking values of order $J$ for the bound states and much larger values for the scattering states.

The model described above could be further extended in interesting ways, some of which we consider in the final section.  But the structure we have built up so far is sufficient to capture the essence of magnetron dynamics, as we discuss next.

\section{Emission from a Fock state}
\label{sec:numerical_results}

On the face of it, the effective quantum model is very different from the classical one and rather more complicated.\footnote{In a certain sense, the quantum model is actually much simpler.  While the classical model is defined in terms of partial differential equations on a continuous space, the quantum model is described by ordinary differential equations on a discrete space.}  This suggests two questions: does the quantum model agree with the classical one?  And does it go beyond it, predicting any new effects?  To answer them, we simulate the model using the quantum jump (Monte Carlo wave function) method~\cite{Dalibard1992, Plenio1998}.

\begin{figure}
\begin{center}
\includegraphics[width=\columnwidth]{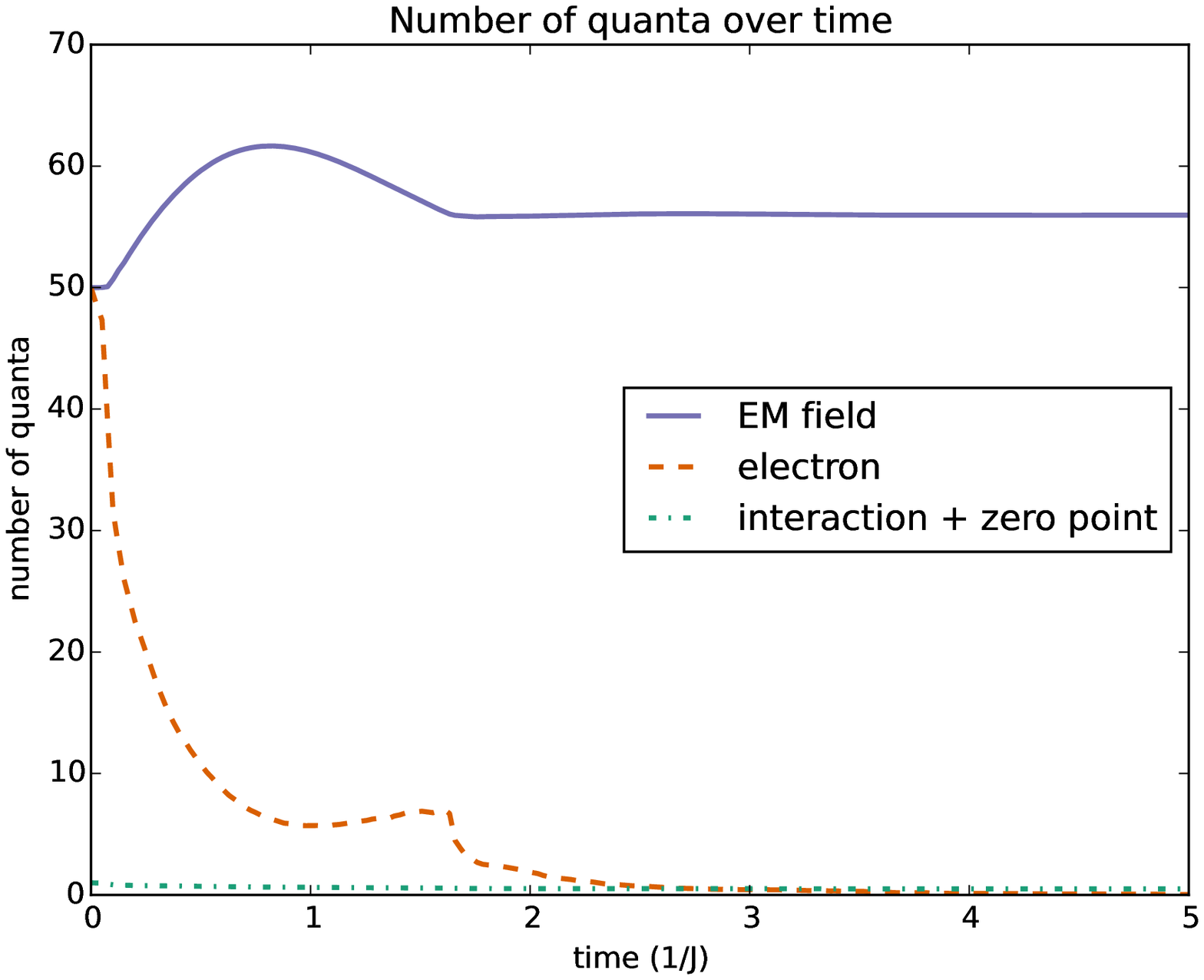}\\
\includegraphics[width=\columnwidth]{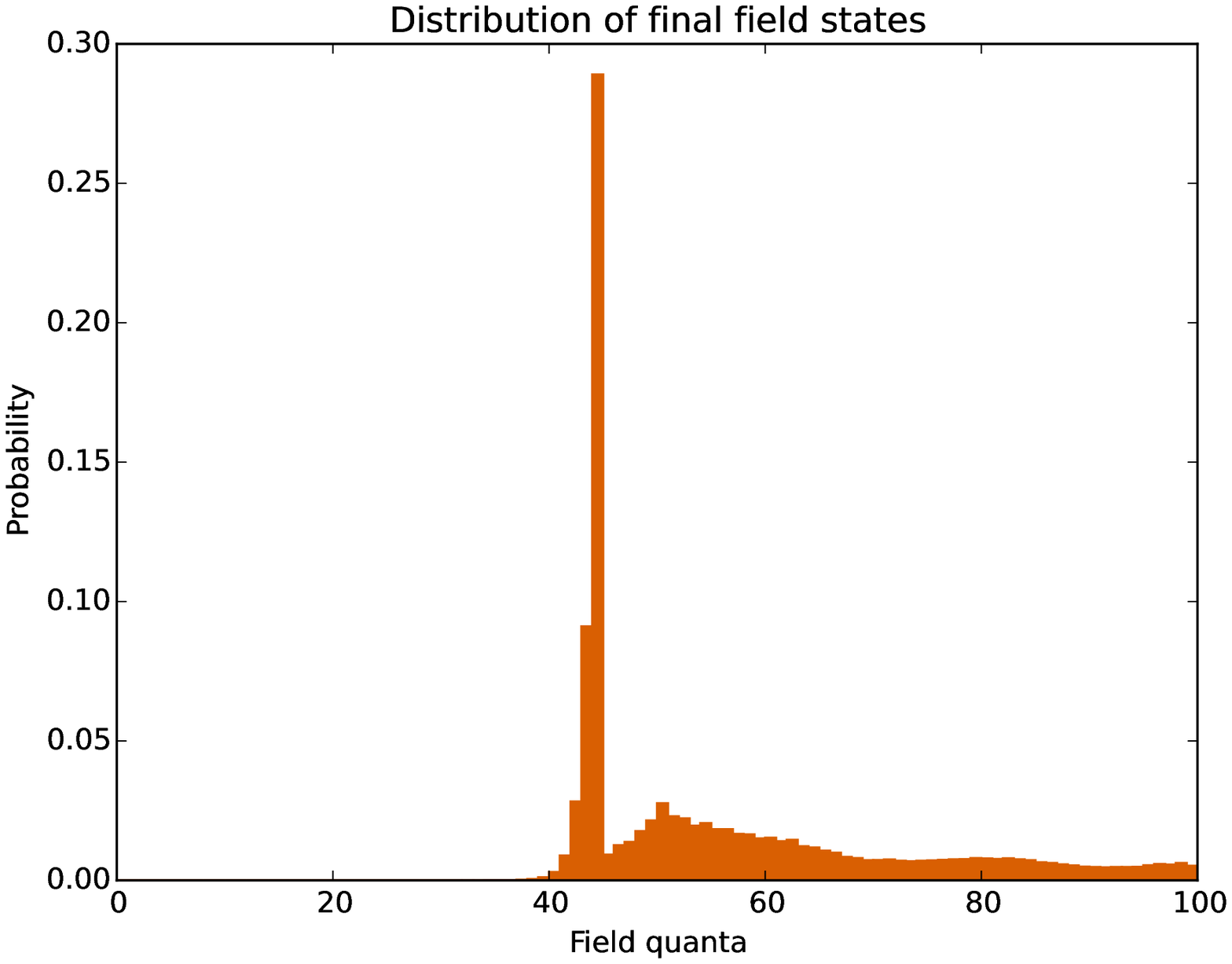}
\caption{Transfer of energy from the DC to the AC field in the quantum magnetron.  The top panel shows the time evolution of the expected number of electron quanta, $\langle \hat{a}^\dagger \hat{a}\rangle$ (red), and AC mode photons, $\langle \hat{b}^\dagger \hat{b}\rangle$ (blue), over time.  The electron decays from the device, but deposits in the AC field some of its energy, which was ultimately derived from the accelerating DC voltage.  The bottom panel shows the final distribution over Fock states of the EM field: an electron could absorb or emit photons, but sequences of more than a few absorptions did not occur.  See the text for further discussion. \label{fig:gain}}
\end{center}
\end{figure}

The central prediction of the classical model is that energy will be transferred on average from the DC electrical field which accelerates the electron to the AC field.  This phenomenon is reproduced in the quantum model.  The top panel of Figure~\ref{fig:gain} shows the expected number of quanta (or energy in units of $\hbar \omega$, or---in the case of the EM mode---number of photons) attributable to the electron and the field over time, for an initial Fock state of the field and the electron in which the two have equal energy.  Since the electron decays from the device, in the long-time limit it contributes nothing to the system's energy.  The field, however, contains more photons at long times than it contained initially.

What is the mechanism behind this process?  The interaction between the electron and the AC field enables emission and absorption events.  Because the Hamiltonian of Eq.~(\ref{eq:eff_hamiltonian_total}) is symmetric with respect to a relabeling of the modes ($\hat{a} \to \hat{b}$, $\hat{b} \to \hat{a}$), an isolated system would, over times $> 1/J$, be equally likely to emit $m$ photons (transfer $m$ quanta from the electron to the field) as to absorb $m$ photons, for any $m$.  In the open system, this symmetry is broken by the decay rates $\{\gamma_n\}$ that depend on $n$, the index of the electron level.  A sequence of many emissions is now more likely than a sequence of many absorptions, because just a few net absorptions will place the electron in a scattering state---the electron will be removed from the device before it can absorb further.  This early termination of chains of net absorptions is illustrated in the bottom panel of Figure~\ref{fig:gain}, which shows the probability distribution over final Fock states of the EM field.  The distribution is dramatically skewed to the right: sequences of many emissions (field quanta $\gg 50$) are common, but those of many absorptions (field quanta $\ll 50$) are never observed. 

\begin{figure}
\begin{center}
\includegraphics[width=\columnwidth]{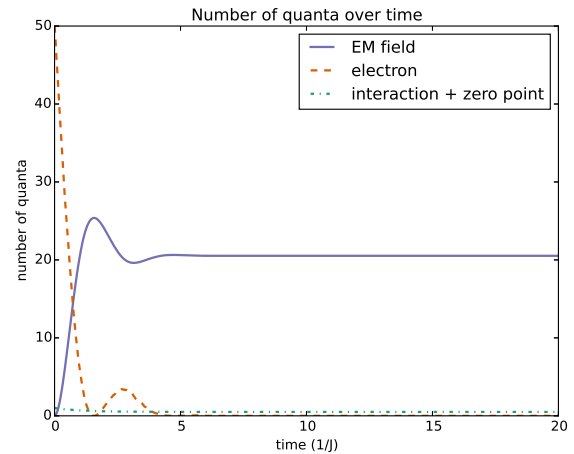}
\caption{The quantum model of the magnetron predicts spontaneous emission.  The expected energy of the electron (red) and field (blue) are plotted over time, for an initial Fock state of the electron and vacuum state of the field.  Initially, the field carries no energy, but by the time the electron has decayed, about half of the electron's initial energy has been transferred to the field.  (The rest of the electron's energy has been lost to the reservoir.)
\label{fig:spontaneous_emission}}
\end{center}
\end{figure}

Amplification of an existing AC field, then, is predicted by both the classical and the quantum models.  In the former model, the device can \emph{only} operate as an amplifier: if the initial amplitude of the AC field is zero, the classical prediction is that it will stay zero.  In contrast, the quantum model of the previous section predicts spontaneous emission even if the field is initially in a vacuum state.  This is illustrated in Figure~\ref{fig:spontaneous_emission}. (In practice, there is always nonzero field present in the device due to thermal fluctuations, and the magnetron will start up without an external input even in the classical model.  At low temperatures, however, the contribution of spontaneous emission should become significant.)




\section{Conclusions and Outlook}

Inspired by the classical model of the cylindrical diode magnetron, we have proposed an effective quantum model consisting of two bosonic modes coupled to a fermionic reservoir.  The quantum model captures the essential behavior of the established classical approach by predicting a net energy transfer from the DC to the AC field.  But the quantum framework can explain a greater range of phenomena, such as spontaneous emission when the field starts out in the vacuum state.

Our results suggest that a solid-state analog of the magnetron would continue to act as a radiation source, with the critical difference that the emission frequency could be elevated to the terahertz range by using a material with small effective mass.

The work discussed here can be extended in interesting ways.  Accurate numerical simulations accounting for the device geometry, perhaps using non-equilibrium Green's function methods~\cite{Datta2000}, are the next natural step. Investigating the wide variety of vacuum tube designs beyond the cylindrical anode magnetron (and the interplay between ballistic electron dynamics and electrodynamics they exploit) is another possibility.  Finally, one could develop entirely novel designs based on materials with nonparabolic band structures such as graphene.  The use of unevenly spaced Landau levels as gain media for lasers had been patented in the 1960s~\cite{Wolff1966}, but at the time thought impossible to realize.  Today we possess both the experimental and theoretical tools to finally implement such concepts.

\begin{acknowledgments}
We wish to thank David Bishop, Joseph Checkelsky, Michael Kolodrubetz, Roberto Paiella, and Stephen W.~Teitsworth for helpful discussions.  TP is grateful for the hospitality of Joshua E.~S.~Socolar, the Duke University Physics Department, and the Graphene Research Centre at the National University of Singapore.

This work was supported in part by Boston University and by Banco Santander.
\end{acknowledgments}

\appendix

\section{Equations of motion for the classical model}
\label{sec:classical_model_derivation}

This section discusses the equations of motion for the classical model used to generate Figure~\ref{fig:magnetron_schematic} and estimate the parameters of the design of Section~\ref{sec:proposed_device}.  The results summarized here are originally due to Hull~\cite{Hull1928}; see Ref.~\cite[Chapter 1]{Ma2004} or~\cite[Chapter 17]{Gilmour2011} for more contemporary discussions.

\paragraph{Equations of motion}

Consider an electron moving in the coaxial electrode arrangement of Figure~\ref{fig:magnetron_schematic}, with its position given in the usual cylindrical coordinates,
\begin{equation*}
\vect{r} = s\, \vect{\hat{s}} + \phi\, \vect{\hat{\phi}} + z\, \vect{\hat{z}}.
\end{equation*}
The electron moves under the influence of the Lorentz force,
\begin{equation*}
\vect{F} = -e\vect{v}\times \vect{B} - e\vect{E} = -eB\vect{v}\times\vect{\hat{z}} + eE \vect{\hat{s}}.
\end{equation*}
The motion of the electron is confined to $z = \mathrm{const.}$; we'll take $z = 0$.  Since
\begin{equation*}
\vect{v} = \dot{s} \,\vect{\hat{s}} + s\dot{\phi} \,\vect{\hat{\phi}},
\end{equation*}
we have
\begin{equation*}
\vect{F} = \vect{\hat{s}} \left(eBs\dot{\phi} + eE\right) + \vect{\hat{\phi}}\left(eB\dot{s}\right)
\end{equation*}
and
\begin{equation*}
\vect{a} = \frac{d\vect{v}}{dt} = \vect{\hat{s}}\left(\ddot{s} - s\dot{\phi}^2\right) + \vect{\hat{\phi}}\left(2\dot{s}\dot{\phi} + s\ddot{\phi}\right).
\end{equation*}
The equations of motion are therefore,
\begin{align*}
\vect{\hat{s}}:&\qquad  \ddot{s} - s\dot{\phi}^2 = -\frac{eB}{\mu}s\dot{\phi} + \frac{eE}{\mu},\\
\vect{\hat{\phi}}:&\qquad 2\dot{s}\dot{\phi} + s\ddot{\phi} = \frac{eB}{\mu}\dot{s}.
\end{align*}
To produce the top panels of Figure~\ref{fig:magnetron_schematic}, these equations were solved numerically with the electric field magnitude given by,
\begin{equation*}
\begin{split}
E(s,t) &= E_\mathrm{DC} + E_\mathrm{AC} \\
       &= \frac{V_\mathrm{DC}}{s\ln \frac{s_c}{s_a}} + \frac{V_\mathrm{AC}}{s_a - s_c} \sin (\omega t + \phi).
\end{split}
\end{equation*}
In the absence of an AC field, the conserved electron energy is given by,
\begin{equation*}
\epsilon = \frac{\mu}{2}\left(\dot{s}^2 + s^2 \dot{\phi}^2\right) - e V_\mathrm{DC} \frac{\ln \frac{s_c}{s}}{\ln \frac{s_c}{s_a}}.
\end{equation*}
In the presence of the AC field, $\epsilon$ becomes a function of time.  The difference between $\epsilon(t = 0)$ and $\epsilon(t = \tau)$ is the net energy gained by the AC field in the $t$ interval $[0, \tau]$.

\paragraph{Hull cutoff voltage}

To find the Hull cutoff voltage, rewrite the second equation of motion as,
\begin{gather*}
\frac{1}{s} \frac{d}{dt}(s^2 \dot{\phi}) = \frac{eB}{\mu}\dot{s},\\
\frac{d}{dt} (s^2 \dot{\phi}) = \frac{eB}{2\mu} \frac{d}{dt} s^2.
\end{gather*}
This implies,
\begin{equation*}
s^2\dot{\phi} = \frac{eB}{2\mu}s^2 + C,\quad \frac{dC}{dt} = 0.
\end{equation*}
If the electron starts from rest at the cathode ($\dot{\phi} = 0$, $s = s_c$), then
\begin{equation*}
0 = \frac{eB}{2\mu}s_c^2 + C \quad \Rightarrow \quad C = - \frac{eB}{\mu} s_c^2
\end{equation*}
and so
\begin{equation*}
\dot{\phi} = \frac{eB}{2\mu}\left(1 - \frac{s_c^2}{s^2}\right).
\end{equation*}
We're interested in trajectories in which the electron barely grazes the anode.  At the apex of such a trajectory, $s = s_a$ and the velocity is purely tangential, so that conservation of energy gives,
\begin{equation*}
\frac{1}{2}\mu \dot{\phi}^2 s_a^2 = eV.
\end{equation*}
Using the $\dot{\phi}$ equation and rearranging, we obtain the Hull voltage condition,
\begin{equation*}
\frac{V_H}{B^2} = \frac{e s_a^2}{8\mu} \left(1 - \frac{s_c^2}{s_a^2}\right)^2.
\end{equation*}

\section{Derivation of the master equation}
\label{sec:master_eq_derivation}

This section presents details of the derivation of Eq.~(\ref{eq:lindblad}).

In the Born approximation, the density matrix of the system in the interaction picture evolves according to~\cite[p.~126]{Breuer2002},
\begin{equation*}
\frac{d\rho}{dt} = - \mathrm{Tr_E}\, \int_0^t\,dt' [\hat{V}_I(t), [\hat{V}_I(t'), \rho(t')\otimes \rho_E(0)]],
\end{equation*}
where $\rho_E(0)$ is the initial density matrix of the environment and $\hat{V}_I$ is the interaction picture coupling, related to the Schr\"{o}dinger picture coupling $\hat{V}$ of Eq.~(\ref{eq:coupling}) by,
\begin{equation*}
\begin{split}
\hat{V}_I &= e^{-\imath (\hat{H}_\mathrm{eff} + \hat{H}_\mathrm{E}) t} \hat{V} e^{\imath (\hat{H}_\mathrm{eff} + \hat{H}_\mathrm{E}) t}\\
 &= \sum_n \sum_\mu \gamma_{n \mu} e^{\imath \hat{H}_\mathrm{eff} t}(\hat{c}_n + \hat{c}_n^\dagger) e^{-\imath \hat{H}_\mathrm{eff} t} \times \\
 &\qquad\qquad e^{\imath \hat{H}_\mathrm{E} t}(\hat{r}_\mu + \hat{r}_\mu^\dagger) e^{-\imath \hat{H}_\mathrm{E} t}.
\end{split}
\end{equation*}
Our first step will be to rewrite this operator in a simpler form.

\subsection{Coupling in the Interaction Picture}

For the environment operators, the anticommutation relations imply,
\begin{equation*}
[\imath \hat{H}_\mathrm{E} t, \, \hat{r}_\mu] = -\imath t \varepsilon_\mu \hat{r}_\mu
\end{equation*}
and so by the Trotter formula,
\begin{equation}
\label{eq:born_approximation}
\begin{split}
\hat{V}_I = \sum_n \sum_\mu &\gamma_{n \mu} e^{\imath \hat{H}_\mathrm{eff} t}(\hat{c}_n + \hat{c}_n^\dagger) e^{-\imath \hat{H}_\mathrm{eff} t}\times \\
&\left(\hat{r}_\mu e^{-\imath \varepsilon_\mu t} + \hat{r}_\mu^\dagger e^{\imath \varepsilon_\mu t}\right).
\end{split}
\end{equation}
We would like to similarly replace the system operator exponentials with phases, but the commutator $[\hat{H}_\mathrm{eff},\, \hat{c}_n]$ is not simply proportional to $\hat{c}_n$, so this is not possible.  Instead, we will rewrite $\hat{V}_I$ in terms of dressed analogs of the system and bath annihilation operators.

Let $\ket{j}$ be an eigenstate of the system Hamiltonian $\hat{H}_\mathrm{eff}$.  We choose the system eigenstates to be simultaneous eigenstates of the electron number operator $\sum_n \hat{c}_n \hat{c}_n^\dagger$.  A resolution of the identity for the Hilbert space of the system is $\sum_j \ket{j}\bra{j} = 1$.  Inserting it twice, after the system time evolution operators,
\begin{equation*}
\begin{split}
\hat{V}_I = \sum_n \sum_\mu \sum_{j,k} &\gamma_{n \mu} X_{jk}^{(n)} \ket{j}\bra{k}\times \\
 &\left(\hat{r}_\mu e^{-\imath \varepsilon_\mu t} + \hat{r}_\mu^\dagger e^{\imath \varepsilon_\mu t}\right) e^{\imath \Delta_{jk} t}
\end{split}
\end{equation*}
where
\begin{gather*}
\Delta_{jk} = \melement{j}{\hat{H}_\mathrm{eff}}{j} - \melement{k}{\hat{H}_\mathrm{eff}}{k},\\
X_{jk}^{(n)} = \melement{j}{\hat{c}_n + \hat{c}_n^\dagger}{k}.
\end{gather*}
Since the system eigenstates were chosen to be simultaneous eigenstates of the electron number operator, $X_{jj}^{(n)} = 0$ for all $j,\,n$.  We split the sum over $k$ into two parts, with a view towards performing a RWA in the system-bath coupling:
\begin{equation*}
\begin{split}
\hat{V}_I &= \sum_{n,\mu,j} \sum_{k: k > j} \gamma_{n \mu} X_{jk}^{(n)} \ket{j}\bra{k} \times \\
 &\qquad\left(\hat{r}_\mu e^{-\imath \varepsilon_\mu t} + \hat{r}_mu^\dagger e^{\imath \varepsilon_\mu t}\right) e^{\imath \Delta_{jk} t} + \\
&\quad \sum_{n,\mu,j} \sum_{k: k < j} \gamma_{n \mu} X_{jk}^{(n)} \ket{j}\bra{k} \times \\
 &\qquad \left(\hat{r}_\mu e^{-\imath \varepsilon_\mu t} + \hat{r}_mu^\dagger e^{\imath \varepsilon_\mu t}\right) e^{\imath \Delta_{jk} t}.
\end{split}
\end{equation*}
Note that,
\begin{equation*}
\begin{split}
\sum_j \sum_{k: k<j} X_{jk}^{(n)} &\ket{j}\bra{k} e^{\imath \Delta_{jk} t} =\sum_k \sum_{j: j<k} X_{kj}^{(n)} \ket{k}\bra{j} e^{-\imath \Delta_{jk} t}\\
 &= \sum_j \sum_{k: k>j} X_{kj}^{(n)} \ket{k}\bra{j} e^{-\imath \Delta_{jk} t}\\
 &=
\sum_j \sum_{k: k>j} X_{kj}^{(n)*} (\ket{j}\bra{k})^\dagger e^{-\imath \Delta_{jk} t},
\end{split}
\end{equation*}
so we may write,
\begin{equation*}
\hat{V}_I = \sum_n \left(\hat{B}_n(t) + \hat{B}_n^\dagger (t)\right)\left(\hat{S}_n(t) + \hat{S}_n^\dagger(t)\right),
\end{equation*}
with
\begin{align*}
\hat{B}_n(t) &= \sum_\mu \gamma_{n\mu} \hat{r}_\mu e^{-\imath \varepsilon_\mu t},\\
\hat{S}_n(t) &= \sum_j \sum_{k: k > j} X_{jk}^{(n)} \ket{j}\bra{k} e^{\imath \Delta_{jk} t}.
\end{align*}
If we assume the system eigenstates are indexed in order of increasing electron number, so that $\hat{S}_n(t)$ is indeed a dressed annihilation operator, we may perform a RWA in the system-bath interaction to obtain,
\begin{equation*}
\hat{V}_I \approx \sum_n \hat{B}_n^\dagger (t) \hat{S}_n(t) + \hat{S}_n^\dagger (t) \hat{B}_n(t).
\end{equation*}

\subsection{Commutators of the Coupling}

As we have noted at the beginning of this Appendix, the evolution of the system density matrix is given by the equation,
\begin{equation*}
\frac{d\rho}{dt} = - \mathrm{Tr_E}\, \int_0^t\,dt' [\hat{V}_I(t), [\hat{V}_I(t'), \rho(t')\otimes \rho_E(0)]].
\end{equation*}
We will now evaluate the commutators appearing on its right hand side.  For clarity, we will drop the explicit time dependence of $\hat{S}$ and $\hat{B}$ (which can be inferred from the index: $m\to t$, $n\to t'$),
\begin{equation*}
\begin{split}
[\hat{V}_I(t'), \,\rho\otimes(t') \rho_\mathrm{E}] = &\sum_n \hat{S}_n^\dagger \rho \otimes \hat{B}_n \rho_\mathrm{E} - \rho \hat{S}_n^\dagger \otimes \rho_\mathrm{E} \hat{B}_n \\
&\quad + \hat{S}_n \rho \otimes \hat{B}_n^\dagger \rho_\mathrm{E} - \rho \hat{S}_n \otimes \rho_\mathrm{E} \hat{B}_n^\dagger
\end{split}
\end{equation*}
and
\begin{widetext}
\begin{equation*}
\begin{split}
[\hat{V}_I(t),\, [\hat{V}_I(t'), \,\rho\otimes(t') \rho_\mathrm{E}]] =
 \sum_{m,n} &\hat{S}_m^\dagger \hat{S}_n^\dagger \rho \otimes (\hat{B}_m\hat{B}_n\rho_\mathrm{E} - \hat{B}_n \rho_\mathrm{E} \hat{B}_m) + (\hat{S}_m^\dagger \hat{S}_n^\dagger \rho - \hat{S}_n^\dagger \rho \hat{S}_m^\dagger)\otimes \hat{B}_n \rho_\mathrm{E} \hat{B}_m \\
  -&\hat{S}_m^\dagger \rho \hat{S}_n^\dagger \otimes (\hat{B}_m \rho_\mathrm{E} \hat{B}_n - \rho_\mathrm{R} \hat{B}_n \hat{B}_m) - (\hat{S}_m^\dagger \rho \hat{S}_n^\dagger - \rho \hat{S}_n^\dagger \hat{S}_m^\dagger)\otimes \rho_\mathrm{E} \hat{B}_n \hat{B}_m \\
  +&\hat{S}_m^\dagger \hat{S}_n \rho \otimes (\hat{B}_m \hat{B}_n^\dagger \rho_\mathrm{E} - \hat{B}_n^\dagger \rho_\mathrm{E} \hat{B}_m) + (\hat{S}^\dagger_m \hat{S}_n \rho - \hat{S}_n \rho \hat{S}_m^\dagger) \otimes \hat{B}_n^\dagger \rho_\mathrm{E} \hat{B}_m \\
  -&\hat{S}_m^\dagger \rho \hat{S}_n \otimes (\hat{B}_m \rho_\mathrm{E} \hat{B}_n^\dagger - \rho_\mathrm{E} \hat{B}_n^\dagger \hat{B}_m) - (\hat{S}_m^\dagger \rho \hat{S}_n - \rho \hat{S}_n \hat{S}_m^\dagger)\otimes \rho_\mathrm{E} \hat{B}_n^\dagger \hat{B}_m.
\end{split}
\end{equation*}
Now, assume the reservoir is in a thermal state.  The trace of the first half of the terms is then zero, and 
\begin{equation*}
\begin{split}
\mathrm{Tr_E}\,[\hat{V}_I(t),\, [\hat{V}_I(t'), \,\rho\otimes(t') \rho_\mathrm{E}]] &= \quad \sum_{m,n} \hat{S}_m^\dagger \hat{S}_n\rho \otimes \mathrm{Tr_E}\,(\hat{B}_m \hat{B}_n^\dagger \rho_\mathrm{E}) - \hat{S}_n \rho \hat{S}_m^\dagger \otimes \mathrm{Tr_E}\,(\hat{B}_n^\dagger \rho_\mathrm{E} \hat{B}_m)\\
&\qquad -\hat{S}_m^\dagger \rho \hat{S}_n \otimes \mathrm{Tr_E}\,(\hat{B}_m \rho_\mathrm{E} \hat{B}_n^\dagger) + \rho \hat{S}_n \hat{S}_m^\dagger \otimes \mathrm{Tr_E}\,(\rho_\mathrm{E} \hat{B}_n^\dagger \hat{B}_m)\\
&\qquad + \mathrm{h.c.}
\end{split}
\end{equation*}
The master equation can be written as a sum of four integrals (and their hermitian conjugates),
\begin{equation}
\label{eq:master_eq_corr_functions}
\begin{split}
\dot{\rho} = \sum_{m,n} \int_0^t\,dt' &\quad \hat{S}_m^\dagger(t) \rho(t') \hat{S}_n(t') \langle \hat{B}_n^\dagger(t') \hat{B}_m(t)\rangle - \rho(t') \hat{S}_n(t') \hat{S}_m^\dagger(t) \langle \hat{B}_n^\dagger(t') \hat{B}_m(t)\rangle \\
 &+ \hat{S}_n(t') \rho(t') \hat{S}_m^\dagger(t) \langle \hat{B}_m(t) \hat{B}_n^\dagger(t')\rangle - \hat{S}_m^\dagger(t) \hat{S}_n(t') \rho(t') \langle \hat{B}_m(t) \hat{B}_n^\dagger(t')\rangle\\
 &+\mathrm{h.c.}
\end{split}
\end{equation}
\end{widetext}
where the bath correlation functions are,
\begin{equation*}
\begin{split}
\langle \hat{B}_n^\dagger(t') \hat{B}_m(t)\rangle &= \mathrm{Tr_E}\,(\hat{B}_n^\dagger(t') \hat{B}_m(t) \rho_\mathrm{E})\\
 &=\mathrm{Tr_E}\,\left(\sum_{\mu,\nu} \gamma^*_{n\mu} \gamma_{m\nu} e^{-\imath \varepsilon_\nu t' - \varepsilon_\mu t} \right)\\
 &=\sum_k n_k \gamma_{nk}^* \gamma_{mk} e^{-\imath \varepsilon_k (t-t')}.
\end{split}
\end{equation*}
with $n_k$ is the Fermi factor of bath level $k$ (and $\langle \hat{B}_m(t) \hat{B}_n^\dagger(t') \rangle$ is defined analogously).  Because we are working in two dimensions, the reservoir density of states is a constant, $\sigma(\omega) = \sigma$.  Converting a sum over the levels into an integral over energies,
\begin{align*}
\langle \hat{B}_n^\dagger(t') \hat{B}_m(t) \rangle &= \int\,d\omega \,\sigma \gamma^*_n(\omega) \gamma_m(\omega) e^{-\imath \omega (t-t')} n(\omega),\\
\langle \hat{B}_m(t) \hat{B}_n^\dagger(t') \rangle &= \int\,d\omega \,\sigma \gamma^*_n(\omega) \gamma_m(\omega) e^{-\imath \omega (t-t')} (1-n(\omega)).
\end{align*}

\subsection{Simplifying the Master Equation}

To make further progress, it is necessary to make additional approximations.  Consider the first integral in Eq.~(\ref{eq:master_eq_corr_functions}) (the other integrals can be treated analogously).  If we assume that the bath correlation functions are memoryless (depend on $t$ and $t'$ only through $\tau = t'-t$) and make the Markov approximation,
\begin{equation*}
\begin{split}
I_1 &= \int_0^t\,dt'\, \hat{S}_m^\dagger(t) \rho(t') \hat{S}_n(t') \langle \hat{B}_n^\dagger (t') \hat{B}_m(t)\rangle\\
 &=\int_0^t\,d\tau\, \hat{S}_m^\dagger(t) \rho(t-\tau) \hat{S}_n(t-\tau) \langle \hat{B}_n^\dagger (\tau) \hat{B}_m(0)\rangle\\
 &=\int_0^\infty\,d\tau\, \hat{S}_m^\dagger(t) \rho(t) \hat{S}_n(t-\tau) \langle \hat{B}_n^\dagger (\tau) \hat{B}_m(0)\rangle
\end{split}
\end{equation*}
Expanding the operators $\hat{S}$ and $\hat{B}$,
\begin{equation*}
\begin{split}
I_1 &= \int_0^\infty\,d\tau\, \sum_j \sum_{k:k>j} \sum_l \sum_{p:p>l}\\
&\qquad X_{jk}^{(m)*} X_{lp}^{(n)} \ket{k}\bra{j} \rho \ket{l}\bra{p} e^{\imath (\Delta_{lp} - \Delta_{jk})t}\\
 &\qquad\times \int\,d\omega\, \sigma \gamma_n^*(\omega) \gamma_m(\omega) e^{-\imath(\omega - \Delta_{lp})\tau} n(\omega).
\end{split}
\end{equation*}
The real part of $I_1$ contributes to the decay rate, while the imaginary part is the frequency shift.  We will ignore the frequency shift, and taking advantage of,
\begin{equation*}
\mathrm{Re}\,\int_0^\infty e^{-\imath \omega t} \, dt = \pi \delta(\omega),
\end{equation*}
will write,
\begin{equation*}
\begin{split}
I_1 &= \sum_j \sum_{k:k>j} \sum_l \sum_{p:p>l}\\
 &\qquad X_{jk}^{(m)*} X_{lp}^{(n)} \ket{k}\bra{j}\rho \ket{l}\bra{p} e^{\imath (\Delta_{lp} - \Delta_{jk})t}\\
 &\qquad\times \pi \sigma \gamma_n^*(\Delta_{lp}) \gamma_m(\Delta_{lp}) n(\Delta_{lp}).
\end{split}
\end{equation*}
We will now make a second rotating wave approximation.  The usual way of doing so would be to drop all terms for which $\Delta_{lp} \neq \Delta_{jk}$.  We will make the milder approximation of replacing $\Delta_{lp}$ with $\Delta_{jk}$ in the argument of $\gamma_m$ in the expression above.  This allows us to write,
\begin{equation}
\label{eq:I_1_post_RWA}
\begin{split}
I_1 &= \sum_j \sum_{k:k>j} \sum_l \sum_{p:p>l}\\
&\qquad \frac{1}{2}\,\hat{A}_{jk}^{(m)\dagger} \rho \hat{A}_{lp}^{(n)}\, e^{\imath (\Delta_{lp} - \Delta_{jk})t} \,n(\Delta_{lp}),
\end{split}
\end{equation}
where we have defined
\begin{equation*}
\hat{A}_{jk}^{(m)} = \sqrt{2\pi\sigma} X_{jk}^{(m)} \gamma_m(\Delta_{jk}) \ket{j}\bra{k}.
\end{equation*}
Performing analogous manipulations for the remaining three integrals in Eq.~(\ref{eq:master_eq_corr_functions}), we obtain
\begin{align*}
I_2 &= \sum_j \sum_{k:k>j} \sum_l \sum_{p:p>l} \frac{1}{2} \rho \hat{A}_{lp}^{(n)} \hat{A}_{jk}^{(m)\dagger} e^{\imath (\Delta_{lp} - \Delta_{jk})t} n(\Delta_{lp}),\\
I_3 &= \sum_j \sum_{k:k>j} \sum_l \sum_{p:p>l} \frac{1}{2} \hat{A}_{lp}^{(n)} \rho \hat{A}_{jk}^{(m)\dagger} e^{\imath (\Delta_{lp} - \Delta_{jk})t} (1- n(\Delta_{lp})),\\
I_4 &= \sum_j \sum_{k:k>j} \sum_l \sum_{p:p>l} \frac{1}{2} \hat{A}_{jk}^{(m)\dagger} \hat{A}_{lp}^{(n)} \rho e^{\imath (\Delta_{lp} - \Delta_{jk})t} (1- n(\Delta_{lp})).
\end{align*}
Now, assume that the anode only absorbs (never emits) electrons.  This means the relevant energy levels of the anode are unoccupied, or $n(\Delta_{lp}) = 0$ for all $l$, $p$.  The master equation is then,
\begin{equation*}
\begin{split}
\dot{\rho} &= \sum_{m,n} \sum_j \sum_{k:k>j} \sum_l \sum_{p:p>l} \\
&\qquad \frac{1}{2} e^{\imath (\Delta_{lp} - \Delta_{jk})t} (\hat{A}_{lp}^{(n)} \rho \hat{A}_{jk}^{(m)\dagger} - \hat{A}_{jk}^{(m)\dagger} \hat{A}_{lp}^{(n)} \rho) + \mathrm{h.c.}
\end{split}
\end{equation*}
Recall that this is the equation for the density matrix in the interaction picture.  The Sch\"{o}dinger picture evolution of the density matrix is given by,
\begin{equation}
\dot{\rho}_S = -\imath [\hat{H}_\mathrm{eff}, \rho_S] + e^{-\imath \hat{H}_\mathrm{eff} t} \dot{\rho} e^{\imath \hat{H}_\mathrm{eff} t}.
\end{equation}
Since the sums over the states in the $\dot{\rho}$ equations above are over eigenstates of $\hat{H}_\mathrm{eff}$, the exponential factors cancel:
\begin{equation*}
\begin{split}
\dot{\rho}_S &= -\imath [\hat{H}_\mathrm{eff}, \rho_S]\\
&\quad + \sum_{m,n} \sum_j \sum_{k:k>j} \sum_l \sum_{p:p>l} \\
&\quad\qquad \frac{1}{2} (\hat{A}_{lp}^{(n)} \rho_S \hat{A}_{jk}^{(m)\dagger} - \hat{A}_{jk}^{(m)\dagger} \hat{A}_{lp}^{(n)} \rho_S)\\
&\quad + \mathrm{h.c.}
\end{split}
\end{equation*}
Let,
\begin{equation*}
\hat{A} = \sum_n \sum_j \sum_{k:k>j} \hat{A}_{jk}^{(n)}.
\end{equation*}
In terms of this operator,
\begin{equation*}
\dot{\rho}_S = -\imath [\hat{H}_\mathrm{eff}, \rho_S] + \hat{A}\rho_S \hat{A}^\dagger - \frac{1}{2} \left(\hat{A}^\dagger \hat{A} \rho_S + \rho_S \hat{A}^\dagger \hat{A}\right),
\end{equation*}
which is Eq.~(\ref{eq:lindblad}).  Let us examine the operator $\hat{A}$:
\begin{equation*}
\hat{A} = \sum_{n=0}^\infty \sum_j \sum_{k: k>j} \sqrt{2\pi\sigma} \melement{j}{\hat{c}_n + \hat{c}_n^\dagger}{k} \gamma(\Delta_{jk}) \ket{j}\bra{k}.
\end{equation*}
We have ordered the eigenstates by their electron number, so that in this sum $\ket{k}$ is always a state with at least as many electrons as $\ket{j}$.  Consequently, $\melement{j}{\hat{c}_n^\dagger}{k} = 0$, and
\begin{equation*}
\hat{A} = \sqrt{2\pi \sigma} \sum_{n=0}^\infty \sum_j \sum_{k: k > j} \gamma_n(\Delta_{jk}) \ket{j}\bra{j} \hat{c}_n \ket{k}\bra{k}.
\end{equation*}
If we strengthen our second RWA [recall the discussion preceding Eq.~\ref{eq:I_1_post_RWA}] by assuming not merely $\gamma_n(\Delta_{lp}) \approx \gamma_n(\Delta_{jk})$, but that $\gamma_n$ is independent of energy, then this expression can be simplified to,
\begin{equation*}
\hat{A} = \sqrt{2\pi\sigma} \sum_{n=0}^\infty \gamma_n \hat{c}_n,
\end{equation*}
which is Eq.~(\ref{eq:lindblad_operator}).

\bibliography{references}

\begin{thebibliography}{29}%
\makeatletter
\providecommand \@ifxundefined [1]{%
 \@ifx{#1\undefined}
}%
\providecommand \@ifnum [1]{%
 \ifnum #1\expandafter \@firstoftwo
 \else \expandafter \@secondoftwo
 \fi
}%
\providecommand \@ifx [1]{%
 \ifx #1\expandafter \@firstoftwo
 \else \expandafter \@secondoftwo
 \fi
}%
\providecommand \natexlab [1]{#1}%
\providecommand \enquote  [1]{``#1''}%
\providecommand \bibnamefont  [1]{#1}%
\providecommand \bibfnamefont [1]{#1}%
\providecommand \citenamefont [1]{#1}%
\providecommand \href@noop [0]{\@secondoftwo}%
\providecommand \href [0]{\begingroup \@sanitize@url \@href}%
\providecommand \@href[1]{\@@startlink{#1}\@@href}%
\providecommand \@@href[1]{\endgroup#1\@@endlink}%
\providecommand \@sanitize@url [0]{\catcode `\\12\catcode `\$12\catcode
  `\&12\catcode `\#12\catcode `\^12\catcode `\_12\catcode `\%12\relax}%
\providecommand \@@startlink[1]{}%
\providecommand \@@endlink[0]{}%
\providecommand \url  [0]{\begingroup\@sanitize@url \@url }%
\providecommand \@url [1]{\endgroup\@href {#1}{\urlprefix }}%
\providecommand \urlprefix  [0]{URL }%
\providecommand \Eprint [0]{\href }%
\providecommand \doibase [0]{http://dx.doi.org/}%
\providecommand \selectlanguage [0]{\@gobble}%
\providecommand \bibinfo  [0]{\@secondoftwo}%
\providecommand \bibfield  [0]{\@secondoftwo}%
\providecommand \translation [1]{[#1]}%
\providecommand \BibitemOpen [0]{}%
\providecommand \bibitemStop [0]{}%
\providecommand \bibitemNoStop [0]{.\EOS\space}%
\providecommand \EOS [0]{\spacefactor3000\relax}%
\providecommand \BibitemShut  [1]{\csname bibitem#1\endcsname}%
\let\auto@bib@innerbib\@empty
\bibitem [{\citenamefont {Liang}\ \emph {et~al.}(2007)\citenamefont {Liang},
  \citenamefont {Neophytou}, \citenamefont {Nikonov},\ and\ \citenamefont
  {Lundstrom}}]{Liang2007}%
  \BibitemOpen
  \bibfield  {author} {\bibinfo {author} {\bibfnamefont {G.}~\bibnamefont
  {Liang}}, \bibinfo {author} {\bibfnamefont {N.}~\bibnamefont {Neophytou}},
  \bibinfo {author} {\bibfnamefont {D.~E.}\ \bibnamefont {Nikonov}}, \ and\
  \bibinfo {author} {\bibfnamefont {M.~S.}\ \bibnamefont {Lundstrom}},\ }\href
  {\doibase 10.1109/TED.2007.891872} {\bibfield  {journal} {\bibinfo  {journal}
  {IEEE Transactions on Electron Devices}\ }\textbf {\bibinfo {volume} {54}},\
  \bibinfo {pages} {677} (\bibinfo {year} {2007})}\BibitemShut {NoStop}%
\bibitem [{\citenamefont {Du}\ \emph {et~al.}(2008)\citenamefont {Du},
  \citenamefont {Skachko}, \citenamefont {Barker},\ and\ \citenamefont
  {Andrei}}]{Du2008b}%
  \BibitemOpen
  \bibfield  {author} {\bibinfo {author} {\bibfnamefont {X.}~\bibnamefont
  {Du}}, \bibinfo {author} {\bibfnamefont {I.}~\bibnamefont {Skachko}},
  \bibinfo {author} {\bibfnamefont {A.}~\bibnamefont {Barker}}, \ and\ \bibinfo
  {author} {\bibfnamefont {E.~Y.}\ \bibnamefont {Andrei}},\ }\href {\doibase
  10.1038/nnano.2008.199} {\bibfield  {journal} {\bibinfo  {journal} {Nature
  Nanotechnology}\ }\textbf {\bibinfo {volume} {3}},\ \bibinfo {pages} {491}
  (\bibinfo {year} {2008})}\BibitemShut {NoStop}%
\bibitem [{\citenamefont {Mayorov}\ \emph {et~al.}(2011)\citenamefont
  {Mayorov}, \citenamefont {Gorbachev}, \citenamefont {Morozov}, \citenamefont
  {Britnell}, \citenamefont {Jalil}, \citenamefont {Ponomarenko}, \citenamefont
  {Blake}, \citenamefont {Novoselov}, \citenamefont {Watanabe}, \citenamefont
  {Taniguchi},\ and\ \citenamefont {Geim}}]{Mayorov2011a}%
  \BibitemOpen
  \bibfield  {author} {\bibinfo {author} {\bibfnamefont {A.~S.}\ \bibnamefont
  {Mayorov}}, \bibinfo {author} {\bibfnamefont {R.~V.}\ \bibnamefont
  {Gorbachev}}, \bibinfo {author} {\bibfnamefont {S.~V.}\ \bibnamefont
  {Morozov}}, \bibinfo {author} {\bibfnamefont {L.}~\bibnamefont {Britnell}},
  \bibinfo {author} {\bibfnamefont {R.}~\bibnamefont {Jalil}}, \bibinfo
  {author} {\bibfnamefont {L.~A.}\ \bibnamefont {Ponomarenko}}, \bibinfo
  {author} {\bibfnamefont {P.}~\bibnamefont {Blake}}, \bibinfo {author}
  {\bibfnamefont {K.~S.}\ \bibnamefont {Novoselov}}, \bibinfo {author}
  {\bibfnamefont {K.}~\bibnamefont {Watanabe}}, \bibinfo {author}
  {\bibfnamefont {T.}~\bibnamefont {Taniguchi}}, \ and\ \bibinfo {author}
  {\bibfnamefont {A.~K.}\ \bibnamefont {Geim}},\ }\href {\doibase
  10.1021/nl200758b} {\bibfield  {journal} {\bibinfo  {journal} {Nano Letters}\
  }\textbf {\bibinfo {volume} {11}},\ \bibinfo {pages} {2396} (\bibinfo {year}
  {2011})},\ \Eprint {http://arxiv.org/abs/1103.4510} {arXiv:1103.4510}
  \BibitemShut {NoStop}%
\bibitem [{\citenamefont {Wang}\ \emph {et~al.}(2015)\citenamefont {Wang},
  \citenamefont {Yang}, \citenamefont {Chen}, \citenamefont {Watanabe},
  \citenamefont {Taniguchi}, \citenamefont {Churchill},\ and\ \citenamefont
  {Jarillo-Herrero}}]{Wang2015}%
  \BibitemOpen
  \bibfield  {author} {\bibinfo {author} {\bibfnamefont {J.~I.-J.}\
  \bibnamefont {Wang}}, \bibinfo {author} {\bibfnamefont {Y.}~\bibnamefont
  {Yang}}, \bibinfo {author} {\bibfnamefont {Y.-A.}\ \bibnamefont {Chen}},
  \bibinfo {author} {\bibfnamefont {K.}~\bibnamefont {Watanabe}}, \bibinfo
  {author} {\bibfnamefont {T.}~\bibnamefont {Taniguchi}}, \bibinfo {author}
  {\bibfnamefont {H.~O.~H.}\ \bibnamefont {Churchill}}, \ and\ \bibinfo
  {author} {\bibfnamefont {P.}~\bibnamefont {Jarillo-Herrero}},\ }\href
  {\doibase 10.1021/nl504750f} {\bibfield  {journal} {\bibinfo  {journal} {Nano
  Letters}\ }\textbf {\bibinfo {volume} {15}},\ \bibinfo {pages} {1898}
  (\bibinfo {year} {2015})}\BibitemShut {NoStop}%
\bibitem [{\citenamefont {Gilmour}(2011)}]{Gilmour2011}%
  \BibitemOpen
  \bibfield  {author} {\bibinfo {author} {\bibfnamefont {A.~S.}\ \bibnamefont
  {Gilmour}},\ }\href@noop {} {\emph {\bibinfo {title} {{Klystrons, Traveling
  Wave Tubes, Magnetrons, Cross-Field Amplifiers, and Gyrotrons}}}}\ (\bibinfo
  {publisher} {Artech House},\ \bibinfo {address} {Norwood, MA},\ \bibinfo
  {year} {2011})\ p.\ \bibinfo {pages} {859}\BibitemShut {NoStop}%
\bibitem [{\citenamefont {Gribnikov}\ \emph {et~al.}(2003)\citenamefont
  {Gribnikov}, \citenamefont {Vagidov}, \citenamefont {Mitin},\ and\
  \citenamefont {Haddad}}]{Gribnikov2003}%
  \BibitemOpen
  \bibfield  {author} {\bibinfo {author} {\bibfnamefont {Z.~S.}\ \bibnamefont
  {Gribnikov}}, \bibinfo {author} {\bibfnamefont {N.~Z.}\ \bibnamefont
  {Vagidov}}, \bibinfo {author} {\bibfnamefont {V.~V.}\ \bibnamefont {Mitin}},
  \ and\ \bibinfo {author} {\bibfnamefont {G.~I.}\ \bibnamefont {Haddad}},\
  }\href {\doibase 10.1063/1.1565496} {\bibfield  {journal} {\bibinfo
  {journal} {Journal of Applied Physics}\ }\textbf {\bibinfo {volume} {93}},\
  \bibinfo {pages} {5435} (\bibinfo {year} {2003})}\BibitemShut {NoStop}%
\bibitem [{\citenamefont {Asada}(2003)}]{Asada2003}%
  \BibitemOpen
  \bibfield  {author} {\bibinfo {author} {\bibfnamefont {M.}~\bibnamefont
  {Asada}},\ }\href {\doibase 10.1103/PhysRevB.67.115303} {\bibfield  {journal}
  {\bibinfo  {journal} {Physical Review B}\ }\textbf {\bibinfo {volume} {67}},\
  \bibinfo {pages} {115303} (\bibinfo {year} {2003})}\BibitemShut {NoStop}%
\bibitem [{\citenamefont {Ryzhii}\ \emph {et~al.}(2009)\citenamefont {Ryzhii},
  \citenamefont {Ryzhii}, \citenamefont {Mitin},\ and\ \citenamefont
  {Shur}}]{Ryzhii2009}%
  \BibitemOpen
  \bibfield  {author} {\bibinfo {author} {\bibfnamefont {V.}~\bibnamefont
  {Ryzhii}}, \bibinfo {author} {\bibfnamefont {M.}~\bibnamefont {Ryzhii}},
  \bibinfo {author} {\bibfnamefont {V.}~\bibnamefont {Mitin}}, \ and\ \bibinfo
  {author} {\bibfnamefont {M.~S.}\ \bibnamefont {Shur}},\ }\href {\doibase
  10.1143/APEX.2.034503} {\bibfield  {journal} {\bibinfo  {journal} {Applied
  Physics Express}\ }\textbf {\bibinfo {volume} {2}},\ \bibinfo {pages}
  {034503} (\bibinfo {year} {2009})}\BibitemShut {NoStop}%
\bibitem [{\citenamefont {Hull}(1928)}]{Hull1928}%
  \BibitemOpen
  \bibfield  {author} {\bibinfo {author} {\bibfnamefont {A.~W.}\ \bibnamefont
  {Hull}},\ }\href {\doibase 10.1109/T-AIEE.1928.5055049} {\bibfield  {journal}
  {\bibinfo  {journal} {Transactions of the American Institute of Electrical
  Engineers}\ }\textbf {\bibinfo {volume} {47}},\ \bibinfo {pages} {753}
  (\bibinfo {year} {1928})}\BibitemShut {NoStop}%
\bibitem [{\citenamefont {Collins}(1964)}]{Collins1964}%
  \BibitemOpen
  \bibinfo {editor} {\bibfnamefont {G.~B.}\ \bibnamefont {Collins}},\ ed.,\
  \href@noop {} {\emph {\bibinfo {title} {{Microwave Magnetrons}}}}\ (\bibinfo
  {publisher} {Boston Technical Publishers},\ \bibinfo {address} {Lexington,
  MA},\ \bibinfo {year} {1964})\ p.\ \bibinfo {pages} {806}\BibitemShut
  {NoStop}%
\bibitem [{\citenamefont {Ma}(2004)}]{Ma2004}%
  \BibitemOpen
  \bibfield  {author} {\bibinfo {author} {\bibfnamefont {L.}~\bibnamefont
  {Ma}},\ }\emph {\bibinfo {title} {{3D Computer Modeling of Magnetrons}}},\
  \href@noop {} {\bibinfo {type} {Phd dissertation}},\ \bibinfo  {school}
  {University of London} (\bibinfo {year} {2004})\BibitemShut {NoStop}%
\bibitem [{\citenamefont {Wickramaratne}, \citenamefont {Zahid},\ and\
  \citenamefont {Lake}(2015)}]{Wickramaratne2015}%
  \BibitemOpen
  \bibfield  {author} {\bibinfo {author} {\bibfnamefont {D.}~\bibnamefont
  {Wickramaratne}}, \bibinfo {author} {\bibfnamefont {F.}~\bibnamefont
  {Zahid}}, \ and\ \bibinfo {author} {\bibfnamefont {R.~K.}\ \bibnamefont
  {Lake}},\ }\href {\doibase 10.1063/1.4928559} {\bibfield  {journal} {\bibinfo
   {journal} {Journal of Applied Physics}\ }\textbf {\bibinfo {volume} {118}},\
  \bibinfo {pages} {075101} (\bibinfo {year} {2015})},\ \Eprint
  {http://arxiv.org/abs/1412.2090} {arXiv:1412.2090} \BibitemShut {NoStop}%
\bibitem [{\citenamefont {Pereira}, \citenamefont {Peeters},\ and\
  \citenamefont {Vasilopoulos}(2007)}]{Pereira2007}%
  \BibitemOpen
  \bibfield  {author} {\bibinfo {author} {\bibfnamefont {J.~M.}\ \bibnamefont
  {Pereira}}, \bibinfo {author} {\bibfnamefont {F.~M.}\ \bibnamefont
  {Peeters}}, \ and\ \bibinfo {author} {\bibfnamefont {P.}~\bibnamefont
  {Vasilopoulos}},\ }\href {\doibase 10.1103/PhysRevB.76.115419} {\bibfield
  {journal} {\bibinfo  {journal} {Physical Review B}\ }\textbf {\bibinfo
  {volume} {76}},\ \bibinfo {pages} {115419} (\bibinfo {year} {2007})},\
  \Eprint {http://arxiv.org/abs/0708.0843} {arXiv:0708.0843} \BibitemShut
  {NoStop}%
\bibitem [{\citenamefont {Zudov}\ \emph {et~al.}(2001)\citenamefont {Zudov},
  \citenamefont {Du}, \citenamefont {Simmons},\ and\ \citenamefont
  {Reno}}]{Zudov2001}%
  \BibitemOpen
  \bibfield  {author} {\bibinfo {author} {\bibfnamefont {M.~A.}\ \bibnamefont
  {Zudov}}, \bibinfo {author} {\bibfnamefont {R.~R.}\ \bibnamefont {Du}},
  \bibinfo {author} {\bibfnamefont {J.~A.}\ \bibnamefont {Simmons}}, \ and\
  \bibinfo {author} {\bibfnamefont {J.~L.}\ \bibnamefont {Reno}},\ }\href
  {\doibase 10.1103/PhysRevB.64.201311} {\bibfield  {journal} {\bibinfo
  {journal} {Phys. Rev. B}\ }\textbf {\bibinfo {volume} {6420}},\ \bibinfo
  {pages} {201311} (\bibinfo {year} {2001})}\BibitemShut {NoStop}%
\bibitem [{\citenamefont {Liu}\ \emph {et~al.}(1988)\citenamefont {Liu},
  \citenamefont {Lin}, \citenamefont {Tsui}, \citenamefont {Lee},\ and\
  \citenamefont {Ackley}}]{Liu1988}%
  \BibitemOpen
  \bibfield  {author} {\bibinfo {author} {\bibfnamefont {C.~T.}\ \bibnamefont
  {Liu}}, \bibinfo {author} {\bibfnamefont {S.~Y.}\ \bibnamefont {Lin}},
  \bibinfo {author} {\bibfnamefont {D.~C.}\ \bibnamefont {Tsui}}, \bibinfo
  {author} {\bibfnamefont {H.}~\bibnamefont {Lee}}, \ and\ \bibinfo {author}
  {\bibfnamefont {D.}~\bibnamefont {Ackley}},\ }\href {\doibase
  10.1063/1.100409} {\bibfield  {journal} {\bibinfo  {journal} {Applied Physics
  Letters}\ }\textbf {\bibinfo {volume} {53}},\ \bibinfo {pages} {2510}
  (\bibinfo {year} {1988})}\BibitemShut {NoStop}%
\bibitem [{\citenamefont {Halperin}(1982)}]{Halperin1982}%
  \BibitemOpen
  \bibfield  {author} {\bibinfo {author} {\bibfnamefont {B.}~\bibnamefont
  {Halperin}},\ }\href {\doibase 10.1103/PhysRevB.25.2185} {\bibfield
  {journal} {\bibinfo  {journal} {Physical Review B}\ }\textbf {\bibinfo
  {volume} {25}},\ \bibinfo {pages} {2185} (\bibinfo {year}
  {1982})}\BibitemShut {NoStop}%
\bibitem [{\citenamefont {Marcuse}(1980)}]{Marcuse1980}%
  \BibitemOpen
  \bibfield  {author} {\bibinfo {author} {\bibfnamefont {D.}~\bibnamefont
  {Marcuse}},\ }\href@noop {} {\emph {\bibinfo {title} {{Principles of Quantum
  Electronics}}}}\ (\bibinfo  {publisher} {Academic Press},\ \bibinfo {address}
  {New York},\ \bibinfo {year} {1980})\ p.\ \bibinfo {pages} {510}\BibitemShut
  {NoStop}%
\bibitem [{\citenamefont {Walls}\ and\ \citenamefont
  {Milburn}(2008)}]{Walls2008}%
  \BibitemOpen
  \bibfield  {author} {\bibinfo {author} {\bibfnamefont {D.~F.}\ \bibnamefont
  {Walls}}\ and\ \bibinfo {author} {\bibfnamefont {G.~J.}\ \bibnamefont
  {Milburn}},\ }\href@noop {} {\emph {\bibinfo {title} {{Quantum Optics}}}},\
  \bibinfo {edition} {2nd}\ ed.\ (\bibinfo  {publisher} {Springer-Verlag},\
  \bibinfo {address} {Berlin, Heidelberg},\ \bibinfo {year} {2008})\ p.\
  \bibinfo {pages} {425}\BibitemShut {NoStop}%
\bibitem [{Note1()}]{Note1}%
  \BibitemOpen
  \bibinfo {note} {Except that the zero-point energy is no longer $1$ but
  rather $\protect \frac {1}{2} + \DOTSB \sum@ \slimits@ _n \protect
  \mathaccentV {hat}05E{c}_n^\dagger \protect \mathaccentV
  {hat}05E{c}_n$.}\BibitemShut {Stop}%
\bibitem [{\citenamefont {Breuer}\ and\ \citenamefont
  {Petruccione}(2002)}]{Breuer2002}%
  \BibitemOpen
  \bibfield  {author} {\bibinfo {author} {\bibfnamefont {H.-P.}\ \bibnamefont
  {Breuer}}\ and\ \bibinfo {author} {\bibfnamefont {F.}~\bibnamefont
  {Petruccione}},\ }\href@noop {} {\emph {\bibinfo {title} {{The Theory of Open
  Quantum Systems}}}}\ (\bibinfo  {publisher} {Oxford University Press},\
  \bibinfo {year} {2002})\BibitemShut {NoStop}%
\bibitem [{\citenamefont {Wiseman}\ and\ \citenamefont
  {Milburn}(2010)}]{Wiseman2010}%
  \BibitemOpen
  \bibfield  {author} {\bibinfo {author} {\bibfnamefont {H.~M.}\ \bibnamefont
  {Wiseman}}\ and\ \bibinfo {author} {\bibfnamefont {G.~J.}\ \bibnamefont
  {Milburn}},\ }\href@noop {} {\emph {\bibinfo {title} {{Quantum Measurement
  and Control}}}}\ (\bibinfo  {publisher} {Cambridge University Press},\
  \bibinfo {year} {2010})\ p.\ \bibinfo {pages} {460}\BibitemShut {NoStop}%
\bibitem [{\citenamefont {Tomka}(2014)}]{Tomka2014}%
  \BibitemOpen
  \bibfield  {author} {\bibinfo {author} {\bibfnamefont {M.}~\bibnamefont
  {Tomka}},\ }\emph {\bibinfo {title} {{Geometric Effects in Quantum
  Non-Equilibrium Dynamics}}},\ \href@noop {} {\bibinfo {type} {Doctoral
  dissertation}},\ \bibinfo  {school} {University of Fribourg} (\bibinfo {year}
  {2014})\BibitemShut {NoStop}%
\bibitem [{\citenamefont {Beaudoin}, \citenamefont {Gambetta},\ and\
  \citenamefont {Blais}(2011)}]{Beaudoin2011}%
  \BibitemOpen
  \bibfield  {author} {\bibinfo {author} {\bibfnamefont {F.}~\bibnamefont
  {Beaudoin}}, \bibinfo {author} {\bibfnamefont {J.~M.}\ \bibnamefont
  {Gambetta}}, \ and\ \bibinfo {author} {\bibfnamefont {A.}~\bibnamefont
  {Blais}},\ }\href {\doibase 10.1103/PhysRevA.84.043832} {\bibfield  {journal}
  {\bibinfo  {journal} {Physical Review A}\ }\textbf {\bibinfo {volume} {84}},\
  \bibinfo {pages} {043832} (\bibinfo {year} {2011})}\BibitemShut {NoStop}%
\bibitem [{Note2()}]{Note2}%
  \BibitemOpen
  \bibinfo {note} {The density of states is a constant because we have assumed
  the device, including the electrodes, to be confined to a plane.}\BibitemShut
  {Stop}%
\bibitem [{Note3()}]{Note3}%
  \BibitemOpen
  \bibinfo {note} {In a certain sense, the quantum model is actually much
  simpler. While the classical model is defined in terms of partial
  differential equations on a continuous space, the quantum model is described
  by ordinary differential equations on a discrete space.}\BibitemShut {Stop}%
\bibitem [{\citenamefont {Dalibard}, \citenamefont {Castin},\ and\
  \citenamefont {Moelmer}(1992)}]{Dalibard1992}%
  \BibitemOpen
  \bibfield  {author} {\bibinfo {author} {\bibfnamefont {J.}~\bibnamefont
  {Dalibard}}, \bibinfo {author} {\bibfnamefont {Y.}~\bibnamefont {Castin}}, \
  and\ \bibinfo {author} {\bibfnamefont {K.}~\bibnamefont {Moelmer}},\ }\href
  {\doibase 10.1103/PhysRevLett.68.580} {\bibfield  {journal} {\bibinfo
  {journal} {Physical Review Letters}\ }\textbf {\bibinfo {volume} {68}},\
  \bibinfo {pages} {580} (\bibinfo {year} {1992})}\BibitemShut {NoStop}%
\bibitem [{\citenamefont {Plenio}\ and\ \citenamefont
  {Knight}(1998)}]{Plenio1998}%
  \BibitemOpen
  \bibfield  {author} {\bibinfo {author} {\bibfnamefont {M.~B.}\ \bibnamefont
  {Plenio}}\ and\ \bibinfo {author} {\bibfnamefont {P.~L.}\ \bibnamefont
  {Knight}},\ }\href {\doibase 10.1103/RevModPhys.70.101} {\bibfield  {journal}
  {\bibinfo  {journal} {Reviews of Modern Physics}\ }\textbf {\bibinfo {volume}
  {70}},\ \bibinfo {pages} {101} (\bibinfo {year} {1998})}\BibitemShut
  {NoStop}%
\bibitem [{\citenamefont {Datta}(2000)}]{Datta2000}%
  \BibitemOpen
  \bibfield  {author} {\bibinfo {author} {\bibfnamefont {S.}~\bibnamefont
  {Datta}},\ }\href {\doibase 10.1006/spmi.2000.0920} {\bibfield  {journal}
  {\bibinfo  {journal} {Superlattices and Microstructures}\ }\textbf {\bibinfo
  {volume} {28}},\ \bibinfo {pages} {253} (\bibinfo {year} {2000})}\BibitemShut
  {NoStop}%
\bibitem [{\citenamefont {Wolff}(1966)}]{Wolff1966}%
  \BibitemOpen
  \bibfield  {author} {\bibinfo {author} {\bibfnamefont {P.~A.}\ \bibnamefont
  {Wolff}},\ }\href@noop {} {\enquote {\bibinfo {title} {Cyclotron resonance
  laser},}\ }\bibinfo {howpublished} {{U.S. Patent No.} 3,265,977} (\bibinfo
  {year} {1966})\BibitemShut {NoStop}%
\end{thebibliography}%
\end{document}